# Adapting CRISP-DM for Idea Mining

A Data Mining Process for Generating Ideas Using a Textual Dataset

Workneh Y. Ayele
Stockholm University
Department of Computer and Systems Sciences, DSV
Stockholm University, Sweden

*Abstract*—Data mining project managers can benefit from using standard data mining process models. The benefits of using standard process models for data mining, such as the de facto and the most popular, Cross-Industry-Standard-Process model for Data Mining (CRISP-DM) are reduced cost and time. Also, standard models facilitate knowledge transfer, reuse of best practices, and minimize knowledge requirements. On the other hand, to unlock the potential of ever-growing textual data such as publications, patents, social media data, and documents of various forms, digital innovation is increasingly needed. Furthermore, the introduction of cutting-edge machine learning tools and techniques enable the elicitation of ideas. The processing of unstructured textual data to generate new and useful ideas is referred to as idea mining. Existing literature about idea mining merely overlooks the utilization of standard data mining process models. Therefore, the purpose of this paper is to propose a reusable model to generate ideas, CRISP-DM, for Idea Mining (CRISP-IM). The design and development of the CRISP-IM are done following the design science approach. The CRISP-IM facilitates idea generation, through the use of Dynamic Topic Modeling (DTM), unsupervised machine learning, and subsequent statistical analysis on a dataset of scholarly articles. The adapted CRISP-IM can be used to guide the process of identifying trends using scholarly literature datasets or temporally organized patent or any other textual dataset of any domain to elicit ideas. The ex-post evaluation of the CRISP-IM is left for future study.

## I. INTRODUCTION

Under the umbrella of data science, the study of extracting value from data has advanced in complexity and size. Data science is more frequently used and is often favored over data mining these days [1]. According to [1], from a metaphorical discourse point of view, if data mining is analogous to gold mining, then data science is comparable with prospecting or searching for profitable mining sites. In this paper, for consistency reasons, data mining is used. Besides, the phrase "data mining" is used in the Cross-Industry-Standard-Process model for Data Mining (CRISP-DM) [2], which has been widely accepted for the past two decades [1]. Data mining is goal-orientated and more focused on processes, while data science is data-oriented and more focused on the exploration of values, including goal-driven values [1]. According to [1], the CRISP-DM process model is developed from goal-oriented perspectives, yet it is still applicable to data science projects.

The CRISP-DM is a generic data mining process model that provides an overview of life cycles of data mining projects [2]. CRISP-DM is popular both in the industry and academia [3], and according to user polls and many surveys, it is considered as the de facto standard for knowledge discovery and data mining projects [1]. The benefits of using the CRISP-DM are reduced cost and time, and minimized knowledge requirements for data mining projects. Moreover, expediting training, knowledge transfer, documentation, and capturing best practices are also the benefits of using CRISP-DM [4] (Chapman et al. 1999). More important, particularly for researchers and practitioners, data mining can be used in innovation endeavors [5]. Hence, CRISP-DM is useful for innovative activities.

Digital innovation is the process in which new or significantly changed artifacts are developed to be embodied in or enabled by IT [6]. In the twenty-first century, data is a goldmine with a potential for stimulating digital innovation. For example, scholarly literature databases are increasingly growing [7]. The accumulation of patent data could also be used to discover innovative solutions, new ideas to existing real-world problems [8]. Solving real-world problems could be supported by generating ideas and stimulating innovation through which economic development could be realized. Economic development arises from people generating ideas [7]. However, it is challenging to analyze large collections of data manually and hence demands innovative ways to deal with it. Another challenge, according to Stevens and Burley, is that initial ideas are seldom commercialized, and it takes thousands of new ideas for a single commercial success [9]. Also, [7] argues that it is becoming harder and harder to elicit innovative ideas. Therefore, it is valuable to have any means to generate as many ideas as possible.

Thorleuchter et al. defined a novel and possibly valuable idea as a text phrase comprising of domain-specific terms from the context of technical language usage rather than everyday language usage [10]. Moreover, a problem-solution pair could also be considered to be an idea [11]. In this paper, "idea" is referred to as: *"text phrase(s), sentence(s) describing new and useful information through expressing possible solution(s) to current problems"*. Ideas mining is used to generate ideas. Thorleuchter et al. define idea mining as *"the use of text mining to automatically process unstructured text data for extracting new and useful ideas"* [10, p.7183]. According to a number of authors, idea mining is introduced by Thorleuchter et al. [12 - 16]. Thorleuchter et al. applied the Euclidean distance measure, which is a distance-based similarity measuring algorithm, to generate ideas [10].

However, distance-based algorithms, which is used to measure the similarity between problem query and solutions from textual datasets, are not the only available solutions to idea mining. For example, deep learning [8], Information Retrieval (IR) [17], topic modeling [18], bibliometric [19], social network analysis [20], association rule mining [21], and collaborative filtering algorithms [11] could also be used for idea mining purposes. Therefore, in this paper, idea mining is described as "*the process of using text mining (e.g., topic modeling) in general, and more specific techniques such as machine learning techniques (e.g., deep learning, association rule mining, collaborative filtering, etc.), social and network theory (e.g., social network analysis), bibliometric, statistical methods, and IR to generate useful and new ideas from unstructured or semi-structured textual data.*"

In addition to distance-based algorithms and text mining techniques presented in the previous paragraphs, trend-based idea elicitation could be applied. Besides, trends relevant to user-demanded products are applicable to generate ideas [22]. Yet, finding ideas from untraceable weak trends that could yield a competitive advantage is like finding a needle in a haystack [23]. Fortunately, machine learning techniques enable the analysis of larger collections of datasets [24]. For example, topic modeling, a type of unsupervised machine learning, can be applied to identify hidden topics [25], and ideas [11]. Also, a topic modeling technique, such as Latent Semantic Analysis (LSA) is used to predict the relevance of ideas and provide valuable insights [26]. According to [27], the elicitation and analysis of trends can be helpful for decision-makers and stakeholders in academia and the industry [27]. Similarly, the elicitation of topics about emerging trends in science and technology is crucial for making decisions [28]. Trend analysis can be used for forecasting trends in technology [29]. Also, forecasting trends in technological advancement could be used to evaluate existing ideas collected or generated by incubators.

The interpretation of results from idea mining techniques needs human intervention. An exemplary scenario from [23] illustrates the relevance of human involvement in idea generation and evaluation. A business analyst of a "digital photography" company found an association between "digital photography" and "mobile phone" 20 years ago and ignored it. The business analyst could have convinced his company's CEOs to introduce a brand new technology, a mobile with an integrated camera. Mobile with an integrated camera is an idea that the business analyst should have proposed, and he most definitely regrets ignoring this idea today. Also, [30] argued that the creative process of generating and evaluating ideas should involve the use of machine learning and human interpretation. Besides, ideas should be evaluated for relevance using quality criteria perspectives such as technical, customer, market, financial, and social. Moreover, each quality criteria consist of concrete evaluation criteria, such as customer perspectives, including novelty, necessity, usability, and usefulness [31].

The use of standard process models for data mining projects contributes to the success of projects. Especially when experienced data scientists do the first projects, then projects will be efficiently, and reliability repeated [2]. Data mining projects rely on standard models for the success of involved people, especially with little time, limited options to experiment with different approaches, and lower technical skills [2]. Existing process models of idea mining mainly use a handful of techniques, namely similarity measure, association rules, and co-occurrence of terms analysis. Moreover, existing process models for idea mining are depicted as simple workflows and pictorial illustrations of processes. On the other hand, the CRISP-DM process model is independent of data mining technology and industry sectors [2]. As a result, it can be adapted for a given scenario. For example, Asamoah and Sharda applied design science research to adapt CRISP-DM for processing social media data to elicit analytical insights in healthcare [32]. Similarly, Spruit and Lytras aligned CRISP-DM's process and design science cycle by linking phases from the two methodologies [33]. CRISP-DM is widely accepted, and it is the de facto standard in data analytics applications [1, 33, 34].

The purpose of this paper is to introduce a reusable process model for idea generation based on CRISP-DM. This study aims to identify relevant phases and corresponding activities by designing and documenting a reusable process model for idea generation using Dynamic Topic Modeling (DTM) and subsequent processes. Therefore, in this study, an extension of a more general task and goal-oriented adaptation of the data mining process model based on CRISP-DM [2] for Idea Mining (CRISP-IM) is proposed. The idea mining processes and the detailed description of the activities could be adapted by documenting each phase followed under CRISP-DM. Yet, in this paper, the design science research approach is applied for the rigor of the study, and to address the "why" and the "how" questions for justifying the components of the proposed artifact. Design science is a relevant artifact development research approach used in information systems, computer science, and engineering research projects [35]. The adapted CRISP-IM model could be used for guiding idea generation in innovation projects. The design of the CRISP-IM is demonstrated using a running example of a data mining process, DTM, and subsequent statistical methods for eliciting ideas.

The dataset used, in this study, is a textual corpus of scholarly articles about self-driving cars. Self-driving technology was chosen since it is a hot research and development topic, attracting many stakeholders. For example, the use of a self-driving car has positive impacts on safety and quality of life, fuel reductions, optimum driving, and crash reductions. Also, with regards to urban planning, efficient parking is a characteristic of self-driving vehicles [36]. However, there are growing concerns utilizing self-driving technology such as ethical issues [37], transportation system, affordability, safety, control, and liabilities [36], and the impact of self-driving technologies on urban planning [37].

This paper is structured into six sections. The second section presents related research, and the subsequent sections include approaches used to adapt CRISP-DM, followed by a description and demonstration of the resulting CRISP-IM, a discussion of the result, and conclusions and future research.

## II. RELATED RESEARCH

According to [1], CRISP-DM has been considered as the de facto standard data mining process model for the past two decades. Also, CRISP-DM is currently used in data science applications. For example, the Data Science Trajectory (DST) model proposed by [1], illustrated in Fig. 1, is backward compatible with CRISP-DM and includes new activities that are common in data science projects, namely the exploration of data sources, goals, products, data values, results, and narratives. Moreover, data acquisition, simulation, architecting, and release are data management related activities. Among the six exploration tasks, result exploration is the process of relating results to business goals. The activities illustrated in the DST model are proposed by [1] to serve as templates for planning data science projects where project managers could include or exclude activities in their workflows based on their demands. Hence, Fig. 1 and the discussions in this chapter illustrate the possibility of adapting CRISP-DM.

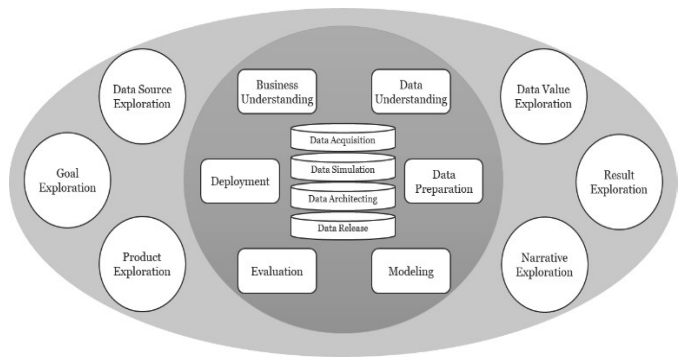

Fig. 1. The Data Science Trajectory (DTS) Map Containing Data Science Exploratory Activities in the Outer Circle, CRISP-DM (Goal-Directed Activities) in the Inner Circle, and Core Data Management Activities at the Center [1, p.5].

In this section, the adaptation of CRSIP-DM in different disciplines, and related research about process models for idea mining, are presented.

### A. Previous Research - Adapting CRISP-DM

The CRISP-DM is adapted in different contexts and disciplines. The following list presents a list of adapted CRISP-DMs.

**Bioinformatics**

1) CRISP-DM is adapted in Bioinformatics for capturing the processes of computational and conventional biological processes of microarray DNA data analysis [38].

**Software Engineering (Software Design)**

2) Atzmueller and Roth-Berghofer extended CRISP-DM by adding explanation dimensions. The explanation dimensions identified are explanation goals (ontological knowledge), kind of explanations (instance knowledge), level of detail (pattern knowledge), and presentation styles (context knowledge) [39].

**Cyber Forensic**

3) Venter et al. adapted CRISP-DM in the cyber forensic domain for Event Mining (CRISP-EM) to define research gaps in evidence mining [40].

**Data Mining Process Model**

4) Martínez-Plumed et al. adapted CRISP-DM by designing a more flexible and Context-aware Standard Process for Data Mining – (CASP-DM), which inherits flexibility and versatility from the CRISP-DM life cycle and put more emphasis in that the sequence of phases is not rigid: context changes may affect different tasks so it should be possible to move to the appropriate phase [41].

**Healthcare**

5) Based on design science research Asamoah and Sharda adapted CRISP-DM for processing big and social network data to generate analytical insights in healthcare [32].

6) Catley et al. adapted CRISP-DM for Temporal Data Mining (CRISP-TDM) by incorporating a multi-dimensional time-series data of medical data streams [42].

7) Adapting CRISP-DM for addressing data mining problems in medicine (CRISP-MED-DM) [43].

8) Spruit and Lytras aligned the design science cycle and CRISP-DM's knowledge discovery process by linking phases from the two methodologies [33].

### B. Previous Research about Idea Mining Models

An idea could be elicited from scholarly literature, patents, reports, the Internet, documents [44], networks of experts [45], social media [46], and crowdsourcing [47]. Furthermore, it is possible to elicit creative ideas of product and service from customers' comments using forums [21].

Previous authors claim that idea mining is introduced by Thorleuchter et al. [10] [12-16]. The idea mining introduced by Thorleuchter et al. uses a text mining technique, a similarity measure using Euclidian distance, to elicit ideas [10]. Similarly, Alksher et al. used similarity measures for the same purpose [44]. However, other idea mining works use different analytics methods. For example, social network analysis [20], deep learning [8], bibliometric [19], topic modeling [18], and IR [17] are used for idea mining.

In spite of the existence of idea mining techniques, the use of standard processes models for idea mining is hardly touched. Existing process models are merely simple flow charts, Business Process Modeling Notations (BPMN) models, or simple nonstandard process diagrams.

1) Simple workflows or simple diagrams as process models: There are five research works presented below, which uses workflows or simple process diagrams to capture idea mining processes.

1) Thorleuchter et al. demonstrated that it is possible to generate ideas using unsupervised machine learning

technique, clustering, to generate ideas. Ideas are generated by measuring the similarity between textual data and problem query using distance metrics, Euclidean distance measures, and finally, applying idea mining measures using Jaccard's coefficient [10]. The process model is illustrated in Fig. 2.

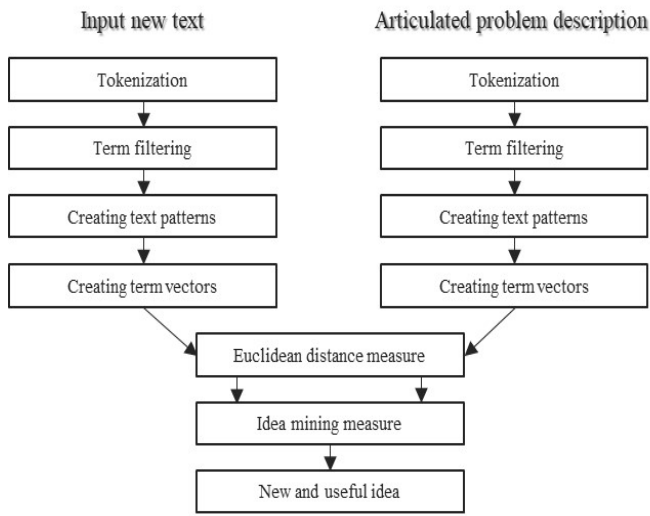

Fig. 2. Thorleuchter Et Al. Proposed a Process Model for Idea Generation using Similarity Measuring Techniques [10, p.7183].

2) It is possible to use the Apriori algorithm for association rule mining by exploring co-occurrence of terms using: key terms repository, associated terms repository, and suggestive terms repository about customer complaint data and online forum data to generate ideas as problem-solution or product-service as an output, the process model is illustrated in Fig. 3. [21].

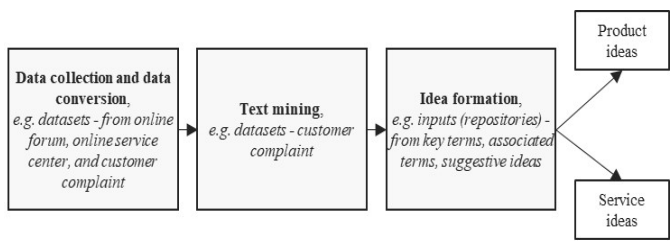

Fig. 3. Kao Et Al. Process Model for Generating Product and Service Ideas [21, p.3].

3) Alksher et al. proposed an idea mining process and framework, as illustrated in Fig. 4. The idea mining uses similarity measures, Euclidean distance measure, to measure the similarity between a term vector of textual data and a new text provided by users [44].

4) Association rule mining and clustering using distance similarity measures are used to generate new ideas, and the process model is illustrated in Fig. 5 below [48].

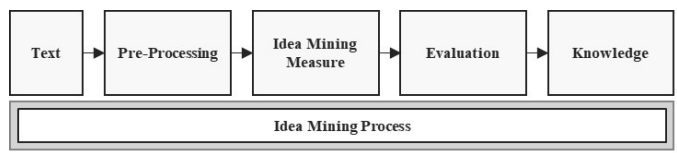

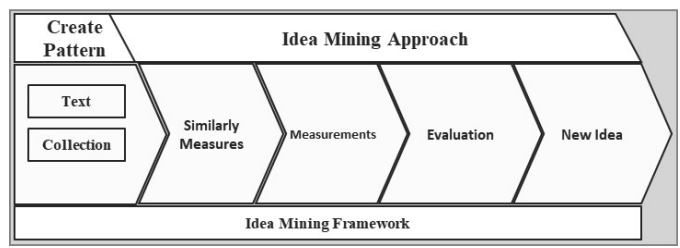

Fig. 4. Alksher et al. Idea Mining Process Model and Framework [44, p.89-90].

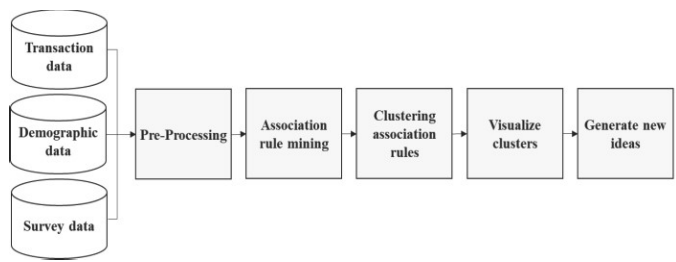

Fig. 5. Idea Mining Process Model using Association Mining [48, p.426].

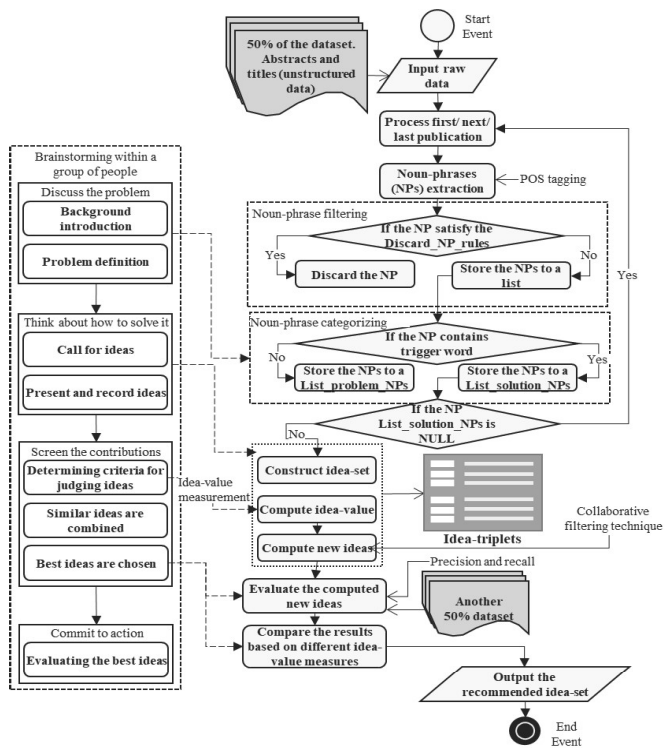

Fig. 6. Liu et al. Idea Mining Process Model using POS and Collaborative Filtering [11, p.6].

5) A set of co-occurring terms, arranged as problem-solution pair is defined as ideas [11]. Liu et al. used Part-Of-Speech (POS) tagging to detect ideas as concepts, noun-phrases. Liu et al. used noun-phrases to classify as a set of problem-solution pairs and added relevance value to making it to triplets. The relevance value is calculated using the idea frequency-inverse document frequency score (phrase co-occurrence). Finally, collaborative filtering is applied on triplets to generate a ranked list of ideas. The process model is illustrated in Fig. 6.

*2) Business Process Modeling Notations:* There is one research work proposing a process model for idea mining presented below.

1) Idea mining process model for generating ideas is proposed using BPMN, see Fig. 7, where textual web data is processed using text mining using Part-Of-Speech tagging, and Natural Language Processing (NLP) such as stemming and lemmatization [46].

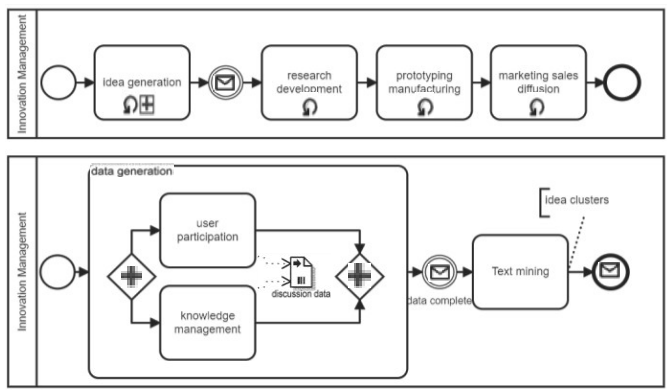

Fig. 7.  Idea Mining Process Model Designed Following Business Process Modeling Notation (BPMN) [46, p.3-4].

## III. Approaches Followed for Adapting CRISP-DM

It is possible to document the use cases, and activities followed while using CRISP-DM to introduce the CRISP-IM for facilitating reusability. Yet, the use of design science, which is an approach used for designing artifacts in computer science, information systems, and related disciplines [35], motivates and justifies as to why the components of the proposed CRISP-IM are relevant. The use of design science also adds value to the rigor of the research. The CRISP-DM for Idea Mining (CRISP-IM) method is adapted after running a DTM that is based on the (Latent Dirichlet Allocation) LDA algorithm. The result of the DTM was also further used to identify predictors and forecast trends using visualization of time-series patterns, and prediction. The development of the artifact, CRISP-IM, was done by following the design science approach, as illustrated in [49].

### A. Dynamic Topic Modeling and Subsequent Analysis

**Dynamic Topic Modeling (DTM):** Large collections of unstructured textual datasets demand machine learning techniques for analyzing and gaining interpretable insights [24]. The DTM model by [50] is based on LDA. LDA is a topic modeling technique designed to elicit hidden topics without temporal information [25]. However, the DTM proposed by Blei and Lafferty could be used to identify topics and their evolutions. Also, to represent topics using multinomial distributions, it uses a state-space-model. Furthermore, it uses Kalman filters vibrational approximation and non-parametric wavelet regression to infer latent topic approximations [50]. In this research, DTM was used to elicit the evolution of topics about self-driving cars. The implementation of the DTM was written in Python, and the subsequent analysis was done using Excel and RStudio.

**Idea generation and evaluation:** The interpretation of the result needs human intervention. For example, the elicitation of ideas is done by examining patterns, trends, and foresight generated from the DTM and the subsequent statistical processing, such as correlation and time series analysis using regression. The use of terms association analysis can be used to make strategic decisions to generate novel product idea development [23]. Topic modeling techniques, such as LDA group co-occurring terms together in respective topics [25]. Elicited topics are labeled with descriptive names, as suggested and illustrated by [25, 51]. Besides, the creative process of generating and evaluating ideas should involve experts and machine learning [30].

### B. Design Science Approach: Method for Designing CRISP-IM

The design science approach, followed in this paper, consists of six activities: the identification of problems, objectives of the solution, design & development, demonstration, evaluation, and communication [49].

*1) Identification of the problem:* Method for problem identification is carried out using a literature review. Existing research on idea generation and evaluation, in particular, the use of DTM and succeeding statistical analysis is hard to find. Furthermore, previous work regarding idea mining, in particular, idea mining through the use of DTMs and succeeding statistical analysis, overlooks the utilization of standard models, such as CRISP-DM.

*2) Objectives of the solution:* The objectives of the artifacts are:

- To be able to follow the guidelines of CRISP-IM to preprocess textual data and identify the best models for topic modeling.

- To be able to follow the guidelines of CRISP-IM to identify topics and their trends.

- To be able to follow the guidelines of CRISP-IM to evaluate the quality of topics identified in terms of interpretability for idea generation and to generate ideas.

*3) Design and development:* Each phase of the CRISP-IM is customized by mapping it with the DTM process carried out to identify emerging trends in [52]. Additionally, the key elements of the phases of the CRISP-IM are inspired by the CRISP-DM [2] model. The detailed specification of components is identified using the DTM, and succeeding statistical analysis processes followed.

The design and development of the CRISP-IM mainly focus on documenting and facilitating the reusability of the innovative process, idea generation. The introduced CRISP-IM guides data analysts to identify trends and generate insightful outputs to generate and evaluate ideas.

*4) Demonstration:* The running example is based on the DTM and the succeeding statistical analysis to demonstrate and motivate parts of the CRISP-IM. Also, the components of each phase of CRISP-IM are motivated from previous research work that involves data collection, preprocessing, selecting the best machine learning model, generating topics, identifying trends, and interpreting the work as presented in [52].

*5) Evaluation and communication:* An empirical evaluation of the CRISP-IM following the design science research approach is left for future study. This paper is written to communicate the result to academia and the industry.

## IV. Result: Adapting CRISP-DM to CRISP-IM

In this section, the adapted CRISP-DM, CRISP-IM, is presented. CRISP-DM to CRISP-IM mapping is illustrated in Table I and Fig. 8. The phases of CRISP-IM are renamed with more relevant and descriptive names, see Table I. Similarly, Venter et al. labeled each phase to suit their process in cyber forensic [40].

### A. Phase 1: Technology Need Assessment

In this phase, a business needs assessment is done through elicitation of business opportunities and challenges being addressed. The inputs in this phase are needs from within a company, and reports from previous idea mining. Also, the identification of goals and success criteria, resources, cost-benefit analysis, risks, and contingencies are carried out.

*1) Motivation for including Phase 1*
*Why is this phase needed?*

Business need assessment is a critical task in data mining projects. Data analytics is used to unlock the hidden potential of large datasets, referred here as idea mining, to produce insights. Despite ever-growing research activities and the resulting findings, it has become difficult to find innovative ideas [7]. The inputs to business need assessment are needs within the organization, goals, success criteria, resources, cost-benefit analysis, and risk and contingencies [2].

*How can the activities in this phase be done?*

Business need assessment can be done through the use of corporate foresight [23] and requirement elicitation. Topic modeling also enables the identification of valuable insights [26]. Forecasting improves creative performance as part of idea generation [53]. Moreover, it is possible to use scientometric, machine learning, and visual analytics to elicit trends and temporal patterns [54].

TABLE I.   Adapting CRISP-DM for idea mining, CRISP-IM

| CRISP-DM | CRISP-IM |
|---|---|
| Business Understanding | Technology Need Assessment |
| Data Understanding | Data Collection and Understanding |
| Data Preparation | Data Preparation |
| Modeling | Modeling for Idea Extraction |
| Evaluation | Evaluation and Idea Extraction |
| Deployment | Reporting Innovative Ideas |

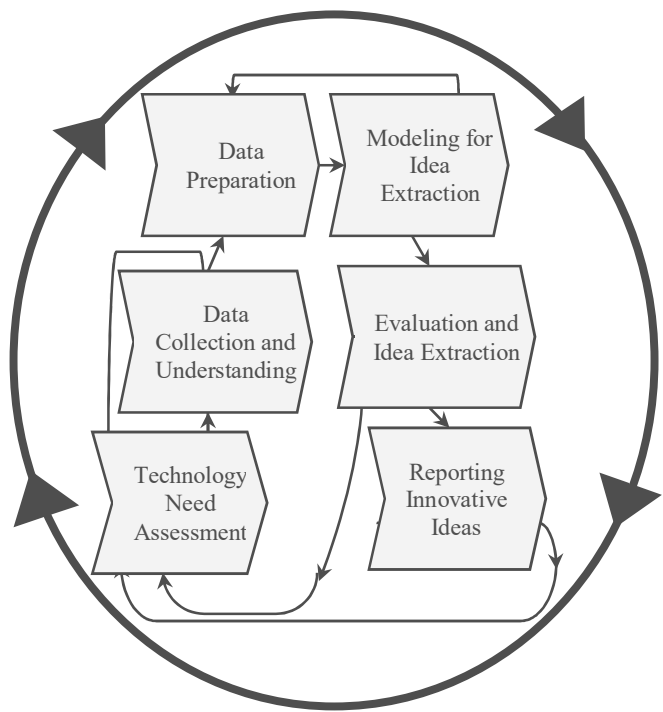

Fig. 8.   CRISP-DM for Idea Mining (CRISP-IM).

*2) Demonstration:* The goal of the running example [52] is to identify emerging trends and patterns to elicit innovative ideas. Hence, [52] identified what could be gained by understanding the project.

The goals of the data mining by [52] are:

- To unlock the potential of ever-growing research findings in academia
- To identify the emerging trends about self-driving cars in academia
- To generate and evaluate ideas for innovation in self-driving cars using scholarly literature as a data source
- To support idea evaluation and generation activities by generating insightful trends

The data source chosen was Scopus, and tools and computing resources used were:

- Python, Excel, a PC with 8GB RAM, a 64 bit Windows 10 Operating System, and an Intel i7 CPU with 2.7 GHz

The analysis of cost-benefit, risk, and contingencies was not done as the purpose of the work by [52] is academic research.

### B. Phase 2: Data Collection and Understanding

In this phase, data is collected after articulating a search query to extract relevant datasets. Data cleaning activity in this phase includes reformatting data into a structured dataset by removing anomalies. Also, exploring and describing data, identification of missing values, removing redundant information, and checking data consistency are also carried out in this phase. Phase 2 could also lead back to Phase 1 – Technology Need Assessment when the data quality is not good enough to extract valid information. For example, when there is insufficient data, we can go back to Phase 1 and include other data sources [2].

#### 1) Motivation for including Phase 2
*Why is this phase needed?*

The most valid activities in this phase are collecting and cleaning the initial data, exploring and describing the data, and finally verifying data quality and adequacy. If the data is not of good quality and is not adequate, go back to Phase 1 to identify other data sources or repositories [2].

*How can the activities in this phase be done?*

Data can be extracted from the chosen data sources, such as Scopus. Scopus has better coverage of journals [55] and contains the latest and larger datasets of scholarly literature than Web of Science [56]. Exploring and understanding data can be carried out using a spreadsheet application [37, 57].

#### 2) Demonstration: Scopus was used as a data source for collecting the dataset. Therefore, a query was formulated to extract the data. A total of 5425 documents were downloaded in CSV format. The documents were retrieved in batches since Scopus limits the maximum downloadable documents to 2000. Removal of duplicates could also be done using reference management software such as Mendeley [58], see Fig. 9. Data understanding was done using Excel and Notepad. A preliminary scanning of the dataset using Excel and Notepad was done to verify data quality. A initial dataset was prepared using Mendeley and Zotero.

### C. Phase 3: Data Preparation - Preprocessing

In this phase, the inclusion of relevant data, cleaning of data, generating derived attributes, merging, and formatting data to make it suitable for modeling are performed.

#### 1) Motivation for including Phase 3
*Why is this phase needed?*

The goal of this phase is to prepare data for modeling. Therefore, the activities performed in this phase address quality issues, identify and apply format for modeling, and clean the data as suggested by [2].

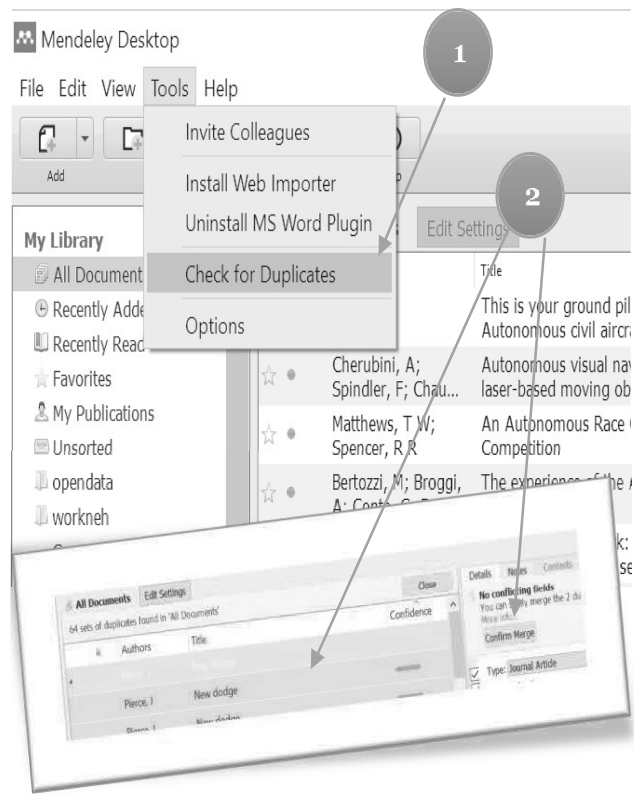

Fig. 9. Screenshot Illustrating how Duplicates are Managed using Mendeley.

*How can the activities in this phase be done?*

Effective text mining processes are predicated on suitable and relevant preprocessing techniques, and preprocessing techniques are applied to unstructured data to generate structured data suitable for text mining models [59].

#### 2) Demonstration: The dataset was preprocessed using Python. The data cleaning activities listed above were done. Preprocessing was aided by word clouds, visualization of term frequency, and list of terms with corresponding frequencies to update the stopword list, as illustrated in Fig. 10. Also, noisy and irrelevant terms such as "IEEE", "Copyright", etcetra were removed. After the tokenization of each document, the lemmatization of terms was done to convert terms to their root forms. Instead of lemmatization, stemming could also be done, as illustrated in [37]. After tokenization, bigrams were computed and added to the tokens list. Finally, the main inputs for the DTM, corpus (bag of words representation), and dictionary were generated. The dictionary representation of the documents was readjusted to contain tokens that are available in more than 100 documents or less than 95% of the documents [52] to minimize the feature dimensionality problem and to include most productive and interpretive data. The combination possibilities and the scale of feature values in text mining are usually greater than standard data mining systems [59].

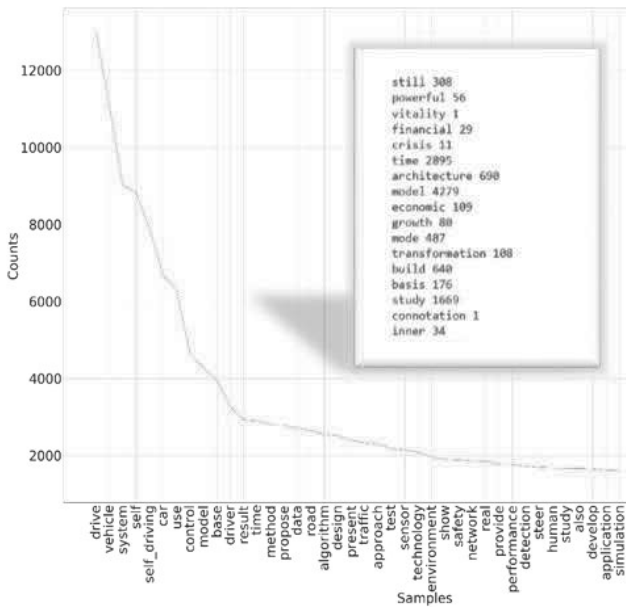

Fig. 10. Word Frequency Visualization and List of Terms with Corresponding Frequencies to Identify Irrelevantly and Misspelled Terms.

*D. Phase 4: Modeling for Idea Extraction*

In this phase, it is possible to do data analysis using tools and techniques such as text mining, network analysis, statistical analysis of linguistic features, etcetera. In this study, the chosen technique is DTM and the succeeding statistical analysis. If the model generated for the DTM is not of good quality, then it is possible to go back to data preparation, Phase 3, if data need to be reformatted to fit specific requirements by the chosen technique.

*1) Motivation for including Phase 4*
*Why is this phase needed?*

In this phase, the most important task is selecting the best model after identifying the data mining modeling technique, building, and assessing it [2]. The quality of the result depends on the quality of the model used.

*How can the activities in this phase be done?*

The inputs to the model generation process are the number of topics, preprocessed text corpus, and dictionary. Perplexity measures are used to determine the number of topics [37] or coherence scores [60]. The best fit model could be generated by comparing the semantic interpretability of models [62].

*2) Demonstration:* The DTM model, which is based on LDA [50], was used to generate topics and their evolution using a dataset about self-driving cars [52]. The DTM proposed by [50], analyzes the evolution of topics in a specific set of chosen datasets. The inputs, corpus, dictionary, and the number of topics are compulsory inputs for running the Python implementation, ldaseqmodel library[1], of the DTM algorithm, which was used in this research. In this phase, models suitable for idea extraction from topic evolution are selected.

---

[1] https://radimrehurek.com/gensim/models/ldaseqmodel.html

1) Selecting the best model for idea extraction. In this paper, DTM based on LDA is chosen. However, DTM topic modeling techniques based on Latent Semantic Analysis (LSA) or Non-negative Matrix Factorization (NMF) could also be chosen based on the generated quality of output, as illustrated in phase 5 by comparing coherence score of these models. LDA-based topic modeling is favored over other topic models such as NMF, and NMF is overlooked [62]. O'callaghan et al. used a coherence score, a technique to measure semantic interpretability of topics, to compare models [62].

- Determine the optimum number of topics - it is possible to determine an optimum number of topics through perplexity [37] or coherence scores [60].

2) Generating the DTM model

3) Generating the output – topics and their evolution identified through visualization and interpretation of results, including succeeding statistical analysis.

- After a preliminary review of terms under each topic, it is possible to find inconsistencies such as the presence of abbreviations or acronyms. Also, the quality of the preprocessed data determines the quality of the model and the result. The optimum number of topics identified could also be affected by the quality of the input corpus. As a result, it is also possible to find overlapping topics despite your efforts in preprocessing and determining the optimum topic number. So you might consider going back to Phase 3 to preprocess accordingly. The quality of the output could also be affected by the NLP, preprocessing, the strategy you are following. For example, if you are choosing either stemming or lemmatization and get poor quality, you can go back to Phase 3, then do a separate preprocessing to compare both strategies.

*E. Phase 5: Evaluation and Idea Extraction*

*1)* The result of Phase 4 is evaluated against the goals of DM listed in Phase 1, and if the result is not in line with the goals of the project, then we need to go back to Phase 1 and redo the whole process again to meet specified goals. The most relevant activities in this phase are listed below.

1) Assessment of model performance

2) Labeling of topics

3) Identifying trends and illustrations using visualizations

4) Prediction of trends through time series analysis when enough time series data is available

5) Run correlation test on candidate terms in the time series to elicit correlations

6) Elicit ideas and align assessment of data mining results with goals and success criteria specified in Phase 1.

- Analyze the result for idea elicitation and evaluation using business analysts and technical experts. In addition, statistical indicators of significance are also used for making Go or No Go decision based on the quality of the result. Furthermore, candidate ideas generated as an output should be evaluated using quality criteria.

- Based on the analysis of the result, it should be possible to make decisions to determine if the result obtained is of acceptable quality and proceed to the next phase by analysts involved. If analysts are not convinced then we need to determine what our next steps are, for example, make an assessment if you need to re-clean the data, by asking questions such as "*Do we need to go back to Phase 1?*", where we repeat the whole processes again from Phase 1- Technology Need Assessment. In Phase 1, you will need to rearticulate goals and success criteria and then continue to Data Collection and Understanding.

*2) Motivation for including Phase 5*
*Why is this phase needed?*

Before generating reports or deploying generated models, it is important to evaluate models [2]. Besides, idea extraction is the main objective of the CRISP-IM model.

*How can the activities in this phase be done?*

**Assessing the model** – Calculation of the coherence score, which measures the quality of topics in terms of coherence of terms within topics, is used for assessing the quality of the model [63].

**Labeling of topics** – The naming of generated topics with descriptive names is not done by machines but by individuals involved in the activity [25, 51].

**Identifying trends and illustrations using visualizations** – Rohrbeck suggested the use of a mixture of people-centric and bibliometric mechanisms to identify weak signals that have potential [23], so it is possible to not miss every possible indicator of trends. Furthermore, text mining techniques are also used for the elicitation of foresight to predict and anticipate the market future, from weak signals [61].

**Predicting trends through time series analysis when enough time series data is available** – Insights and foresight generated using text analytics, topic modeling, can be used to elicit ideas [26]. Similarly, it is possible to elicit trends and identify temporal patterns using machine learning, scientometric, and visual analytics [54]. Furthermore, when there is enough observation to run a time series prediction, then it is possible to run forecasting of trends for a chosen topic. For example, according to [64], if you have observations of at least four, it is arguably advisable to use regression. Besides, [65] suggest that at least 50 observations are required to use advanced models such as Autoregressive Integrated Moving Average (ARIMA) model for time-series predictions.

**Running correlation tests on candidate terms** – Correlation is used to identify relationships or associations between variables [66]. Rohrbeck suggested the identification of association is valuable to identify potential product ideas [23]. Statistical significance tests should be used to determine the quality of post-processing using correlations, regressions, and time-series analysis results. Moreover, weak signals of change in the evolution of trends of topics should not be ignored.

**Eliciting ideas and align assessment of data mining results with goals and success criteria specified in Phase 1** – Elicitation of ideas can be done by using topic modeling [26] and using association of terms found in publications and patents [23]. Evolving trends identified from topics can be as inputs for decision making in research and real-world technology [28]. Analysis of trends results in forecasting trends in technology [29], forecasting while idea generation improves idea evaluation [53].

Idea generation and evaluation activities should include the use of criteria such as customer perspectives – with attributes such as necessity, novelty, usefulness, usability, and other perspectives such as technical, market, financial, and social [31]. In addition to expert judgment to evaluate the quality and acceptability of the result, it is possible to run statistical significance tests on predictions [52].

*3) Demonstration:* Before labeling topics with descriptive names, an overall assessment of the coherence of terms under each topic was done. The calculation of the coherence score was done based on the algorithm by [63], and the implementation was based on the genism [2] python library. Several numbers of topics were assigned to models to choose a model with the highest coherence score, and the coherence score was generated for each number of topics, as illustrated in Fig. 11.

Labeling of topics was done by assigning descriptive names to topics identified, as suggested in [25] and [51]. Besides, thematic analysis of terms under each topic was done to label topics following [67]. However, when acronyms and abbreviations are present, labeling of topics is difficult [68]. Therefore, acronyms and abbreviations were interpreted after identifying their interpretations accordingly. The elicitation of trends was carried out using graphical illustrations of the result, see Fig. 12, for illustration. The visualization of term probabilities throughout the timeline enables the explanation and interpretation of trends [69], see Fig. 12.

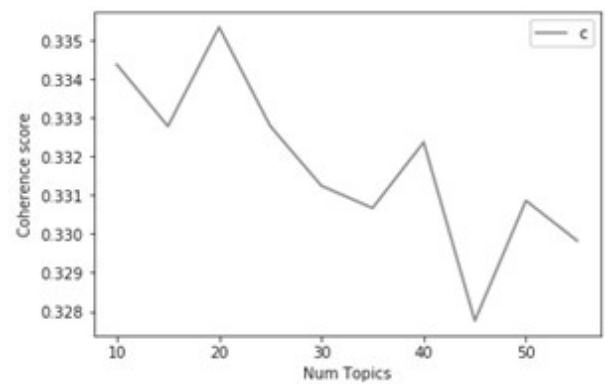

Fig. 11. Topic Coherence Score.

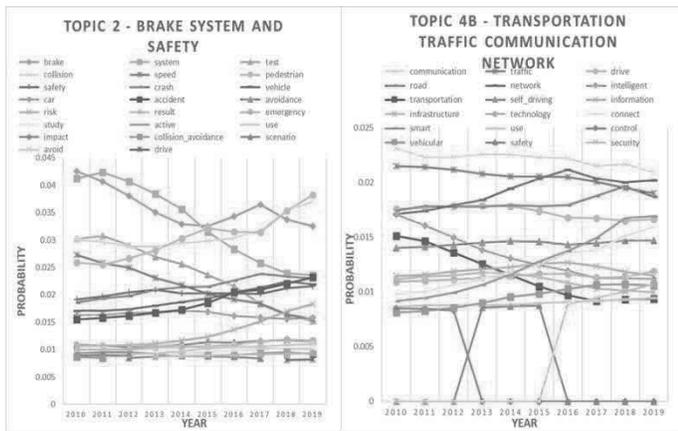

Fig. 12. Illustration of Topic Evolution.

The term "Smart," for example, has an increasing trend as illustrated in Topic 4B, Fig. 12, indicating that there is an increasing interest in academia in the concept smart in relation to terms such as road, traffic, communication, and so on. This trend implies issues regarding smart cities are also gaining attention in academia. On the other hand, Topic 2 shows that pedestrian, collision, accident, and break have increasing trends; also, safety is trending, implying that there is an increased demand for control and safety of self-driving cars. Innovative ideas related to these trends are likely to have a higher relevance. Finally, if there is enough observation of time series data, then it is possible to run a time series analysis for forecasting and to extract valuable insights.

Thematic analysis and visualization were used to elicit and interpret trends. Identified trends can be used to extrapolate and generate ideas and reports. Generated ideas can be used for evaluating the relevance and timeliness of ideas being commercialized by incubators, innovators, and R&Ds. The thematic analysis of evolving topics was done to analyze, elicit trends, and report patterns and themes [67]. It is possible to identify the correlation of terms using scatterplots of a chosen topic, as presented in [52], and it is illustrated in Fig. 13. Also, it is possible to generate innovative product ideas from associated terms [23].

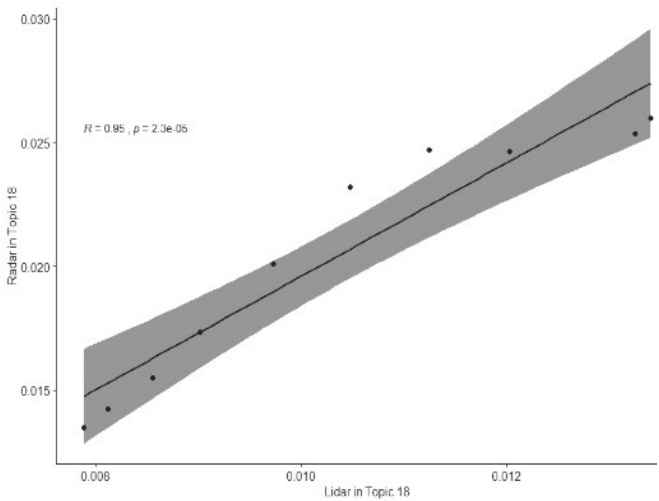

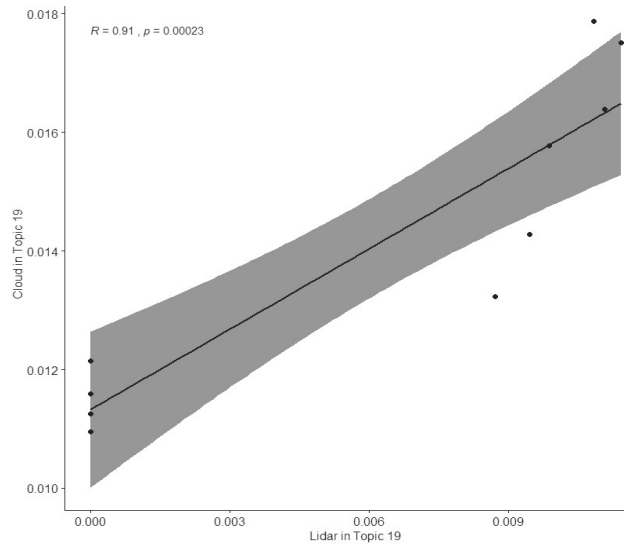

Fig. 13. Illustration of Correlation between Lidar and Radar on the Top and Lidar and Cloud on the Bottom.

In the demonstration [52], there are ten observations, which is a ten years' period. Since the number of observations is ten years' period, it is not advisable to run advanced time series analysis like ARIMA model [65], hence simple regression analysis for time series prediction, as suggested by [64], was done. For the illustration of prediction using regression, see Fig. 14.

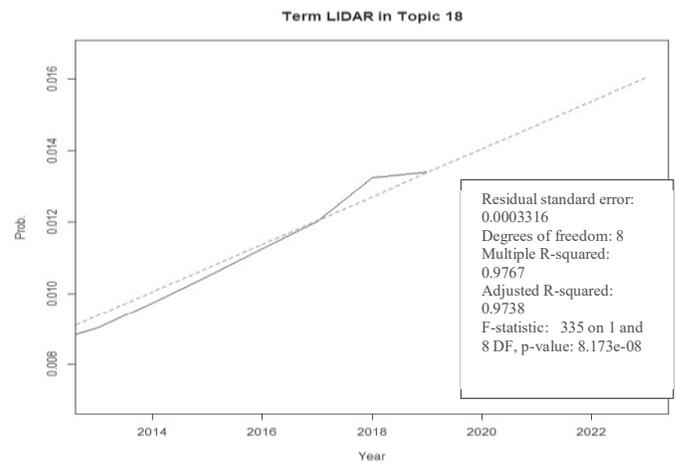

Fig. 14. Illustration of Forecasting Trends, the Choice of the Time-Series Model could be Evaluated using Statistical Significanc Measures such as Residual Standard Error, and P-Values.

Finally, identified innovative lists of ideas and research agendas were documented [52]

### F. Phase 6: Reporting Innovative Ideas

In this last phase, analysis and interpretation of generated ideas and information from the result are reported. The main tasks suggested in this phase are reporting results, documenting best practices, and deployment planning. If the model generated in Phase 5 is of high quality, and if the task, idea generation, is done frequently, then it could be integrated with existing applications, and finally, maintenance and monitoring could be done if the model is integrated with existing applications.

*1) Motivatioan for including Phase 6*
*Why is this phase needed?*

The model alone is not the end product of data mining projects, and hence knowledge elicited need to be articulated and presented in a reusable way so that it can be used by customers for the future. Planning deployment, maintenance, and monitoring, producing, and reviewing final reports are activities in the last phase suggested in CRISP-DM [2].

*How can the activities in this phase be done?*

In this phase, generating reports is done and documenting actions that need to be done to make use of the created models so that it can be reused [2]. The documentation of innovative ideas and best practices can be done using project management tools and word processing applications.

*2) Demonstration:* The reporting of the results is done through textual interpretation and explanation through the visualization of topic evolution. Additionally, lessons learned, implementation codes, use cases, insights leading to quest further, and best practices are documented for future analysis. Finally, ideas for research and commercialization are documented and reported [52].

## V. Discussions

This paper aims to build a set of guidelines for processing scholarly articles to generate and evaluate ideas using machine learning. In addition to machine learning, idea generation can be done using a network of experts [45], social media and online forums [21, 46], crowdsourcing [47], and innovation contests [70-71]. Ideas generated through machine learning can be used as an input in the front end of innovation activities and could ultimately be commercialized. For example, according to [72], an idea "gluten-free-beer" generated from a machine learning extracted from online forums was used by Lakefront Brewery Inc. to introduce the first gluten-free beer. Thus, idea generation activities are beneficial to companies. Idea mining unlocks potential marketing possibilities for organizations. On the other hand, the activities of idea generation, idea mining, can benefit from and following standard process models. Typically, CRISP-DM facilitates reusability, learning, knowledge transfers, cost and time reduction, and documentation [2].

In this paper, CRISP-DM is adapted to capture the processes of DTM for idea mining process, CRISP-IM, by using a dataset of self-driving cars. Similarly, CRISP-DM is adapted in different research domains such as cyber forensic for Event Mining (EM) as CRISP-EM [40], healthcare for processing big social network data [32], and other contexts as illustrated in Sections 2 and 3. Also, standard process models enable people with lower technical skills to easily follow and carry out sophisticated data mining tasks [2]. Likewise, CRISP-IM facilitates the reuse of best practices and expedites successful idea mining tasks. The result of this study could be used as a guideline for processing scientific literature, patents, and any textual information with a temporal variable, for evaluating and eliciting ideas through the analysis of trends.

CRISP-DM does not deal with tasks related to project organization, management, and quality issues [3]. However, the adapted CRISP-IM is designed to facilitate reusability with minimal knowledge requirements, documenting best practices, and research contribution purposes. Therefore, the proposed model does not focus on project planning and organization, similar to other adapted models, for example [40].

## VI. Conclusions And Future Research

The growth of research findings has become exponential, and user-generated textual data is growing at an unprecedented pace while it is hard to find innovative ideas. It is possible to unlock the potential of digital data sources such as patents, social media, crowdsources, and etcetera by applying machine learning. For example, it is possible to identify research agendas and innovative ideas by analyzing research findings. The primary purpose of this research is to facilitate idea generation by simplifying idea mining and unlocking the untapped potential of growing unstructured or semi-structured textual data. To facilitate reusability, knowledge management, ease of learning, are the benefits of using standard process models. The proposed CRISP-IM adds values to the front-end of the innovation processes by streamlining the elicitation of innovative ideas.

Idea mining in this paper was used to demonstrate the process of conducting exploratory data analytics, and machine learning (unsupervised learning) on scholarly articles to elicit new ideas. It is also possible to generate ideas using a variety of other techniques. For example, to generate ideas, the following algorithms and methods could be used: Euclidean distance measure, deep learning, IR, topic modeling, bibliometric, social network analysis, association rule mining, statistical analysis, and collaborative filtering algorithms. Idea mining could be conducted to extract valuable and new information from mainly unstructured textual data. Text mining is the main technique used in idea mining activities, which involves IR and most data mining tasks such as association rule mining, similarity measuring techniques, and topic modeling. Text mining a field of study within computer science that uses techniques of data mining, IR, machine learning, NLP, and knowledge management [59]. Therefore, it is possible to undertake idea mining through the use of techniques of text mining, social network analysis, and bibliometric. Therefore, idea mining activities involve text mining, bibliometrics, statistics, and social network analysis.

Future studies should address the limitation of this study, ex-post evaluation, and extend this study by including other techniques and data sources for idea mining. Future studies could evaluate the applicability of CRISP-IM in different contexts. Finally, it is suggested to extend the CRISP-IM to include other types of DTMs such as Dynamic LSI, and Dynamic NMF.

### Acknowledgment

I wish to thank Alexey Voronov (Ph.D.) and Mahedre DW Amanuel (M.Sc.) at RISE for their support in their inputs at the beginning of the project. I would also like to thank Gustaf

Juell-Skielse (Associate Professor), Paul Johannesson (Professor), Panagiotis Papapetrou (Professor) for offering constructive feedback. This study was conducted as part of the research project IQUAL (2018-04331) funded by Sweden's Innovation Agency (Vinnova).